\documentclass[runningheads]{llncs}
\usepackage{graphicx}
\usepackage{amsmath} 
\usepackage{amssymb} 
\usepackage[hidelinks]{hyperref}
\usepackage{url} 
\usepackage[strings]{underscore}
\usepackage[T1]{fontenc}
\usepackage{cfr-lm}
\usepackage[ttdefault=true]{AnonymousPro}

\begin{document}

\title{Multi-Agent Simulation for AI Behaviour Discovery in Operations Research} 
%
%
\author{Michael Papasimeon\orcidID{0000-0003-1184-8376} \\
Lyndon Benke\orcidID{0000-0003-3088-4129} }
\authorrunning{M. Papasimeon and L. Benke} 
%
\institute{Defence Science and Technology Group\\
506 Lorimer Street, Fishermans Bend, VIC. 3207, Australia\\
\email{{firstname.lastname}@dst.defence.gov.au} }

\newcommand{\acezero}{\texttt{ACE0}}

\maketitle              

\begin{abstract}

    We describe \acezero{}, a lightweight platform for evaluating the suitability and viability of AI methods for
    behaviour discovery in multi-agent simulations. Specifically, \acezero{} was designed to explore AI methods for
    multi-agent simulations used in operations research studies related to new technologies such as autonomous aircraft. 
    Simulation environments used in production are often high-fidelity, complex,
    require significant domain knowledge and as a result have high R\&D costs. 
    Minimal and lightweight simulation environments can help researchers and engineers evaluate the
    viability of new AI technologies for behaviour discovery in a more agile and potentially cost effective manner. 
    In this paper we describe the motivation for the development of \acezero{}.
    We provide a technical overview of the system architecture, describe a case study of behaviour discovery in the aerospace domain, and provide a qualitative evaluation of the system. The evaluation includes a brief description of collaborative research projects with academic partners, exploring different AI behaviour discovery methods.

\end{abstract} 

\section{Introduction} 
\label{sec:intro} 
In this paper we provide an overview of \acezero{}, a lightweight multi-agent-based simulation (MABS) environment designed for evaluating  AI behaviour discovery methods for operations research studies.
In operations research and analysis, multi-agent simulations have a long track record of being used to evaluate technologies for acquisition and their subsequent employment. In the aerospace domain, multi-agent simulations have been used to model, simulate and ultimately compare and assess aircraft to support acquisition programs and to help evaluate how they may be operated at both a tactical and strategic level. In large engineering projects, these constructive simulation environments allow large organisations in government and industry to reduce cost and risk on complex projects.

In many of these simulations, agent behavioural models have been used to represent the decision making of both human and
autonomous systems. For example, a significant body of work exists around using agent models to represent pilot decision making in constructive simulations of air operations~\cite{2008:Heinze:SimPilots,2007:Papasimeon:HAVE,1998:Tidhar,1992:Jones}. Typically, agent oriented software engineering (AOSE) techniques are used to elicit domain knowledge~\cite{2017:Evertsz:CM} and to then handcraft agent behaviour models using technologies such as finite state machines, behaviour trees, or more sophisticated approaches such as the belief-desire-intention (BDI) model of agent reasoning. However, one of the limitations of using traditional AOSE techniques is that the domain knowledge elicited and ultimately programmed in agent code represents current operational practices for existing technologies. The introduction of new technologies such as autonomous aircraft (also commonly known as UAV; unmanned air vehicles) poses a challenge for the development of agent behavioural models, as they are unlikely to be operated in the same way that traditional aircraft are operated. 

Hence, there is a requirement to augment traditional AOSE techniques with exploratory AI methods from fields such as machine learning, evolutionary algorithms and automated planning. The long term goal is to discover novel tactics, strategies and concepts of operations (CONOPS) that current domain experts may otherwise not have considered~\cite{2002:Smith}. 
We use the term \emph{behaviour discovery} to include all of these methods and their application.

One of the challenges in evaluating new exploratory AI methods for their viability for behaviour discovery, is the complexity of production simulation environments. Production simulators are often complex, requiring significant software engineering and domain expertise to deploy effectively, and typically involve the interplay of many high-fidelity computational models. The added complexity of deployment on high performance computing clusters can make it cost prohibitive (in schedule and resources) to use one of these environments to evaluate the viability of an exploratory AI algorithm. Often there are additional complications relating to intellectual property and security that make academic collaboration difficult.

\begin{figure}[t]
    \begin{center} 
    \includegraphics[width=1.0\textwidth]{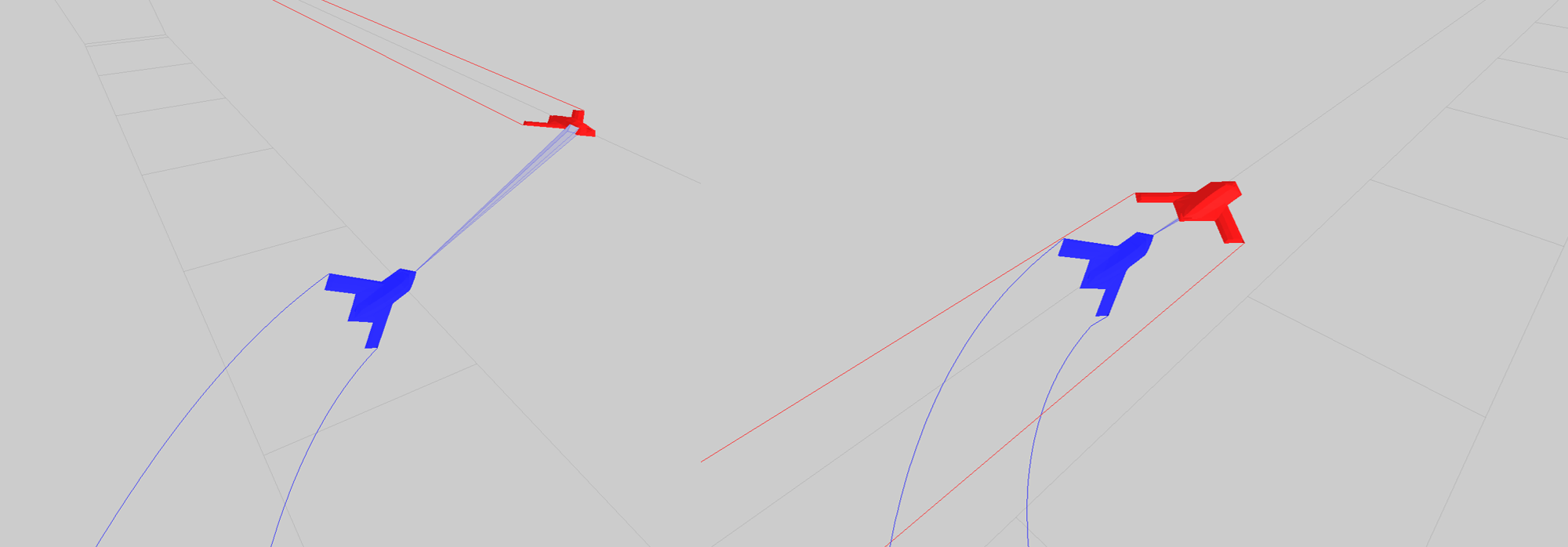}
    \end{center}
    \label{fig:xcombat}
    \caption{3D visualisation of autonomous aircraft simulated in \acezero{}}
\end{figure} 

In this paper we propose using a lighter weight, lower fidelity MABS for evaluating exploratory AI methods that does not incur the overhead of a production environment. We present \acezero{} as one such lighter weight MABS, and describe our experience to date in evaluating exploratory AI methods in the aerospace domain. 

The remainder of the paper provides a high level system overview of the \acezero{} MABS, followed by a case study of its application in the aerospace tactics domain, with an evaluation of some of the exploratory AI methods that have been investigated to date. We conclude by discussing some of the future challenges for agent behaviour discovery with multi-agent simulation.

Our contributions in this paper are threefold; (a) we pose the problem of agent behaviour discovery as future challenge for the fields of both multi-agent simulation and agent oriented software engineering; (b) we present \acezero{} as a reference MABS for conducting exploratory AI analysis in a lightweight environment; and (c) we outline a series of lessons and challenges for agent behaviour discovery arising from our experience with \acezero{} in the aerospace domain.

\section{System Overview}
In this section we provide a high level overview of \acezero{} and its associated components. As mentioned in
Section~\ref{sec:intro}, \acezero{} was developed as a research simulator to reduce the costs of exploring AI algorithms in a production simulation environment. In the air combat operations analysis domain, lightweight research simulators are not uncommon, with LWAC~\cite{2020:Toubman:Thesis} and AFGYM~\cite{2020:RAND:AI} being recent examples. While \acezero{} does have a focus on the aerospace domain, the architecture is generic and can be extended to model naval and ground entities. Furthermore, the architecture supports the grouping of entities into teams, and allows agents to undertake command and control of both individual entities and teams of entities. This allows for the modelling and representation of joint (air, maritime and land) operations. The complexities of team modelling are outside the scope of this paper and hence the focus will be on exploratory AI methods for multiple agents outside of a team structure.

\begin{figure}[t]
    \centering
    \includegraphics[width=0.8\textwidth]{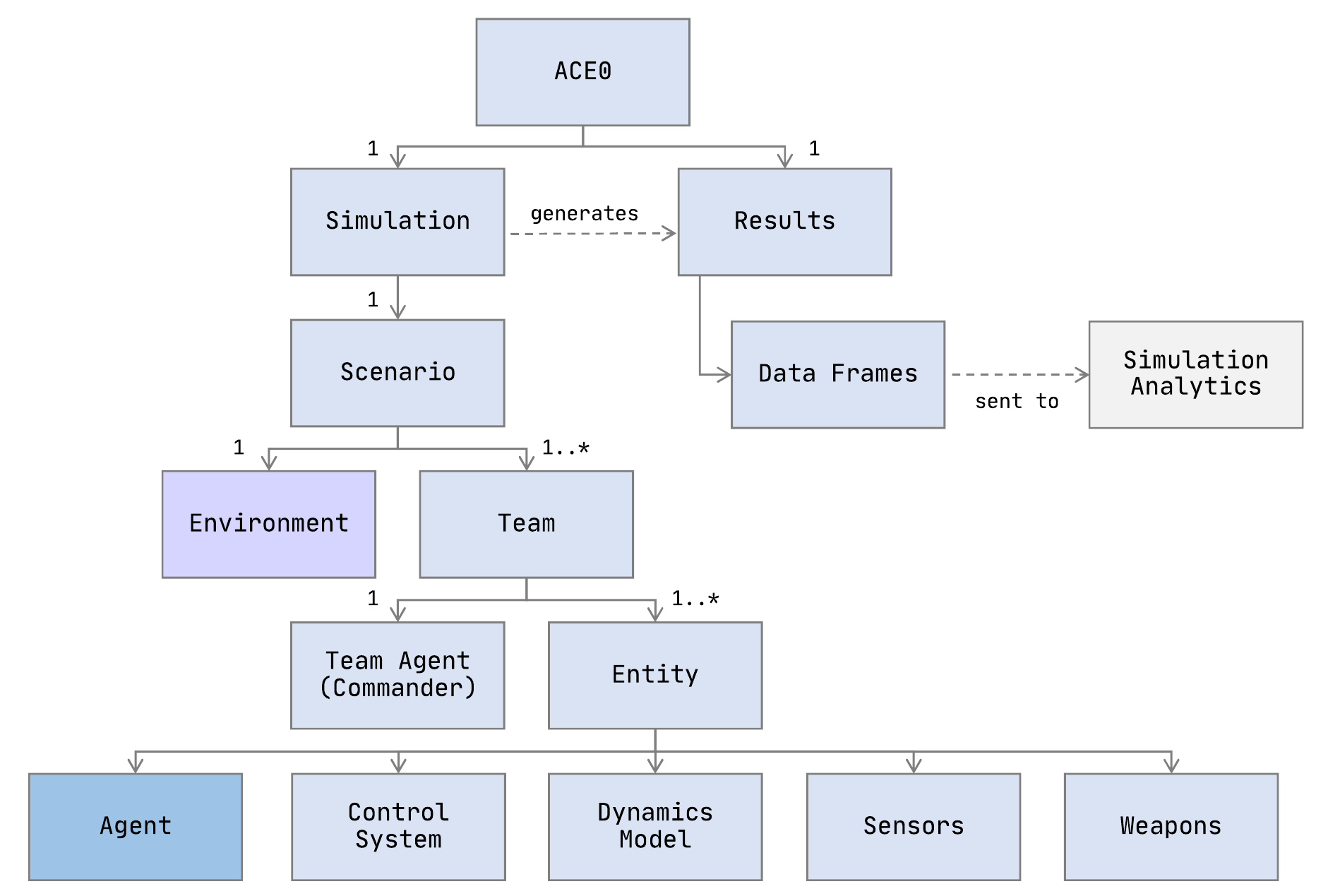}  
    \caption{High level static design architecture of the \acezero{} MABS.}
    \label{fig:ace0design} 
\end{figure}

We begin our description of \acezero{} with a high level UML diagram shown in Figure~\ref{fig:ace0design}. At the highest level \acezero{} consists of a time-stepped multi-agent simulation engine, and a results generator that is used for post simulation run analytics. The top entity being simulated is a \emph{Scenario}, which can be assembled from a user specified library of predefined entities (for example different types of aircraft) and their associated initial conditions. The \emph{Scenario} consists of a model of the \emph{Environment} which facilitates communication and interaction between entities, and one or more teams specified by the \emph{Team} class. Each team is made up a \emph{Team Agent}, typically representing the decision making of the team commander or leader, and one or more \emph{Entity} objects. Each object represents an embodied entity such as an aircraft, ship or vehicle. Each \emph{Entity} is made up of a number of computational components. These include sensor and weapon models, and a dynamics model that represents how the entity moves through the physical environment. The control system is a separate model that can take high level commands from an agent and translate them into lower level commands that are understood by the dynamics model. Finally, the agent model represents the decision making model for the entity, taking as input the entire entity state (dynamics, sensors, weapons etc.) and generating commands for all these components. The \emph{Agent} represents the decision making component of the entity, which may be a model of a human decision making or the reasoning component of an autonomous system.

When representing an autonomous aircraft, the pilot agent can reason about higher level decision making, and generate higher level actions (such as flying to a waypoint or changing heading). The flight control system (FCS) model has the responsibility for translating these into low level aircraft roll, pitch and throttle commands which are understood by the flight dynamics model. As will be discussed later, selecting a suitable level of action abstraction has a significant effect on the suitability and viability of exploratory AI algorithms for behaviour discovery. 

\begin{figure}[t]
    \centering
    \includegraphics[width=1.0\textwidth]{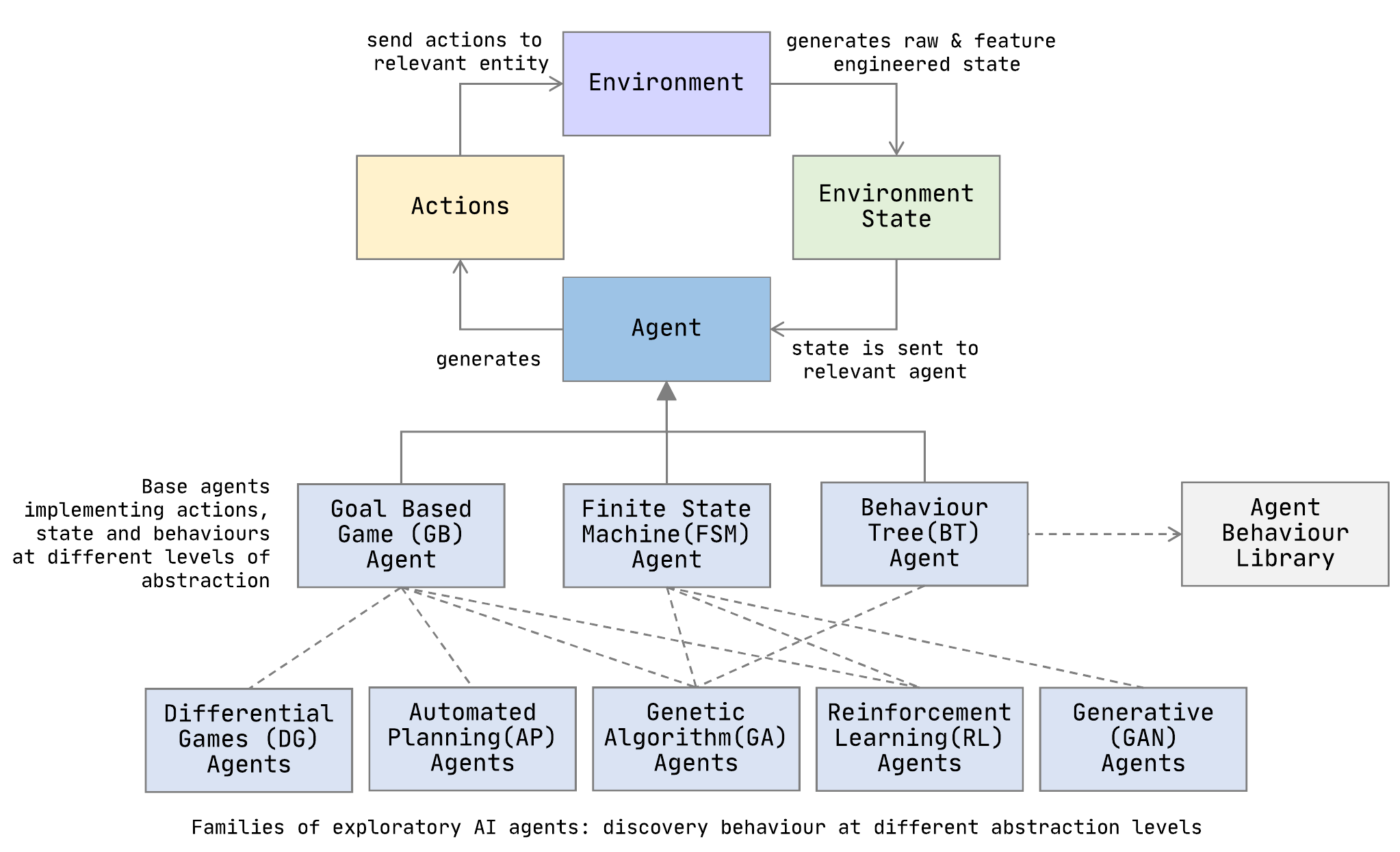}  
    \caption{A spectrum of behaviour abstractions (with respect to agent actions, behaviours and state space representations) are supported in \acezero{}.}
    \label{fig:ace0agents} 
\end{figure}

The ability to reason about behaviour at different levels of abstraction is key to evaluating different exploratory algorithms for behaviour discovery. 
This is critical for client driven operations analysis studies where the level of abstraction selected for the state and action spaces has a significant effect on how the results of the study can be presented and explained. One can imagine that a study investigating behaviour algorithms for a control system would use lower level representations of system state and action when compared to studies which focus on behaviour discovery at the tactical, operational or strategic level.

To address this requirement, \acezero{} supports a spectrum of abstraction levels for behaviour discovery. This is demonstrated in Figure~\ref{fig:ace0agents}.
In the top half of Figure~\ref{fig:ace0agents} we see the common \emph{$\langle Agent,Action,Environment,State \rangle$} reasoning loop. However, the abstraction level for agent action and environment state varies depending on the type of agent being used. \acezero{} provides examples of three levels of agent abstraction, including low-level implementations using goal-based agents, intermediate levels using finite state machines (FSMs), and higher levels where behaviours are assembled into reusable behaviour trees~\cite{2017:Collendanchise:BT,2014:Marzinotto:BT}. This allows the evaluation of an exploratory AI algorithm at different levels of abstraction. Figure~\ref{fig:ace0agents} shows examples of agents integrated into \acezero{} using different approaches. In some cases, such as for evolutionary algorithms, they were evaluated at all three of the levels of abstraction exposed by \acezero{}. Specific details in the context of a use case will be provided in the next section.

\section{Case Study: Aerial Manoeuvring Domain} 
\label{sec:background} 
\begin{figure}[t]
    \centering
    \includegraphics[width=0.6\columnwidth]{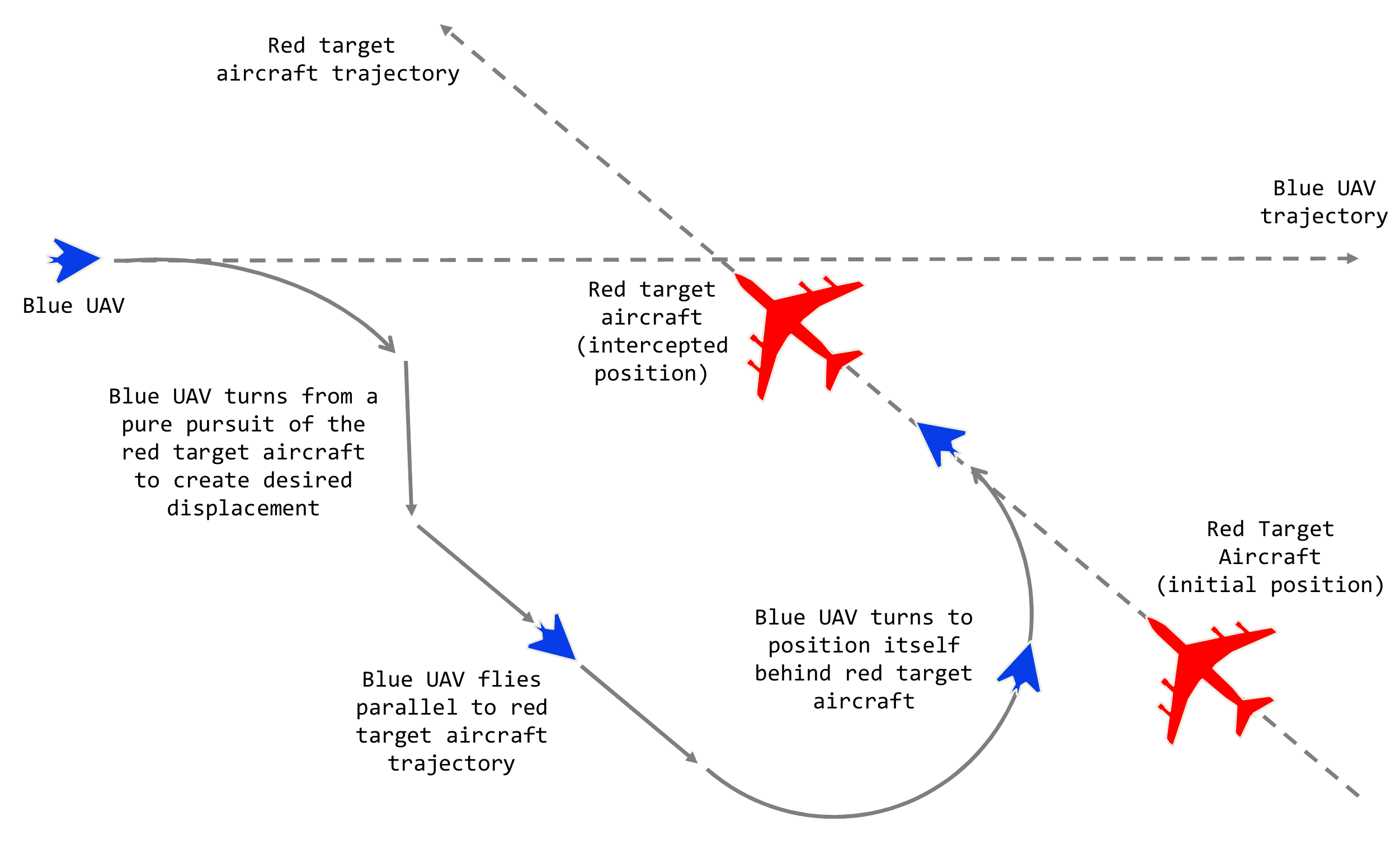} 
    \includegraphics[width=0.3\columnwidth]{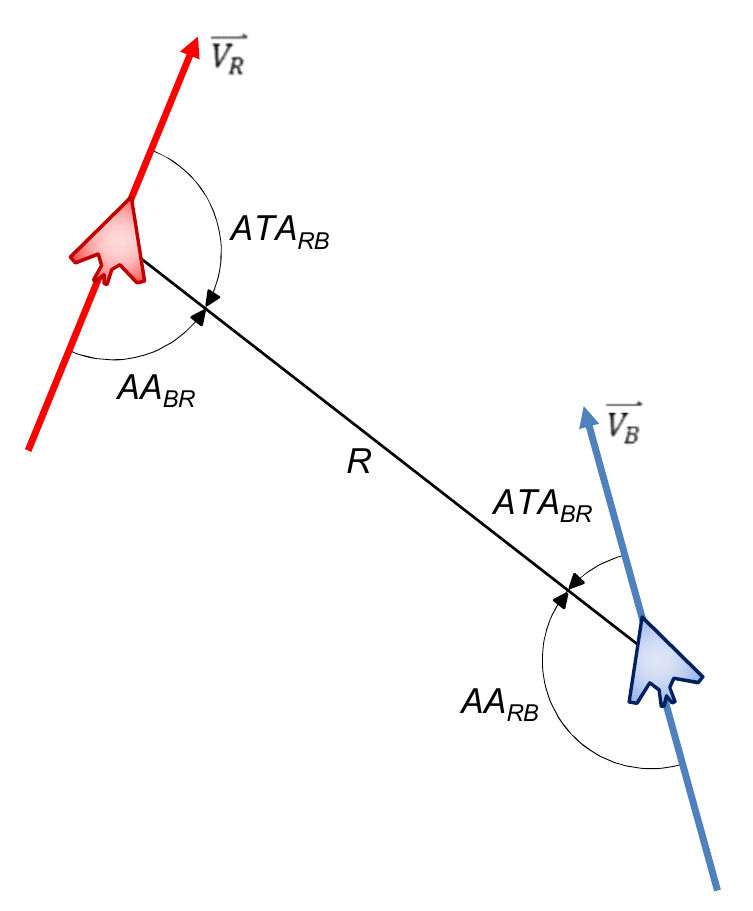}  
    \caption{(Left) Schematic of a stern conversion intercept.
    (Right) Relative Orientation: $AA_{RB} + ATA_{BR} = AA_{BR} + ATA_{RB} = \pi$ 
    where  $ATA \in (-\pi, \pi)$ and $AA \in (-\pi, \pi)$.} 
    \label{fig:relative}  
\end{figure}

In this section we present a case study from the aerial manoeuvring domain, which involves one autonomous aircraft manoeuvring behind another while maintaining the position for a certain amount of time. We use this simple scenario as a case study (a) as a way of providing an initial evaluation of exploratory algorithms, (b) because it requires a relatively simple explanation in terms of domain knowledge, (c) it can be implemented at multiple levels of action abstraction, and (d) it can be easily implemented using well known behaviour specification methods such as finite state machines or behaviour trees, providing standardised baselines.
This type of manoeuvre is called a \emph{Stern Conversion Intercept}~\cite{1985:Shaw} and can be employed operationally for a number of reasons including formation flying, aerial refuelling, visual identification or in the case of air combat situations, weapons employment. A schematic of this manoeuvre can be seeing in Figure~\ref{fig:relative}a.

While it is possible to provide raw environmental state information to various exploratory algorithms (such as the position and orientation of both aircraft), we can consider a smaller subset of features if we consider the relative orientation of the two aircraft.
Consider two aircraft (denoted \emph{blue} and \emph{red}) flying relative to each other at separation distance $R$. We can define their
relative orientation through a number of angles as shown in Figure~\ref{fig:relative}b. From the perspective of the
\emph{blue} aircraft, the \emph{red} aircraft is at an antenna train angle $ATA_{BR}$ relative to the \emph{blue's} aircraft's velocity vector
$\vec{V_B}$.  The \emph{blue} aircraft is also at at an aspect angle of $AA_{BR}$ relative to the \emph{red} aircraft's tail (the
anti-parallel of the red velocity vector $\vec{V_R}$). From the perspective of the \emph{blue} aircraft, we can define any situation~\footnote{This is a simplified view of the situation, as one can consider higher order features such as turn rates and other time derivatives of the basic state space variables.} using four features; $ATA$, $AA$, $R$ and the difference in velocity $\Delta V$. 

\begin{figure*}[t]
    \centering
    \begin{tabular}{cc} 
        \includegraphics[width=0.48\textwidth]{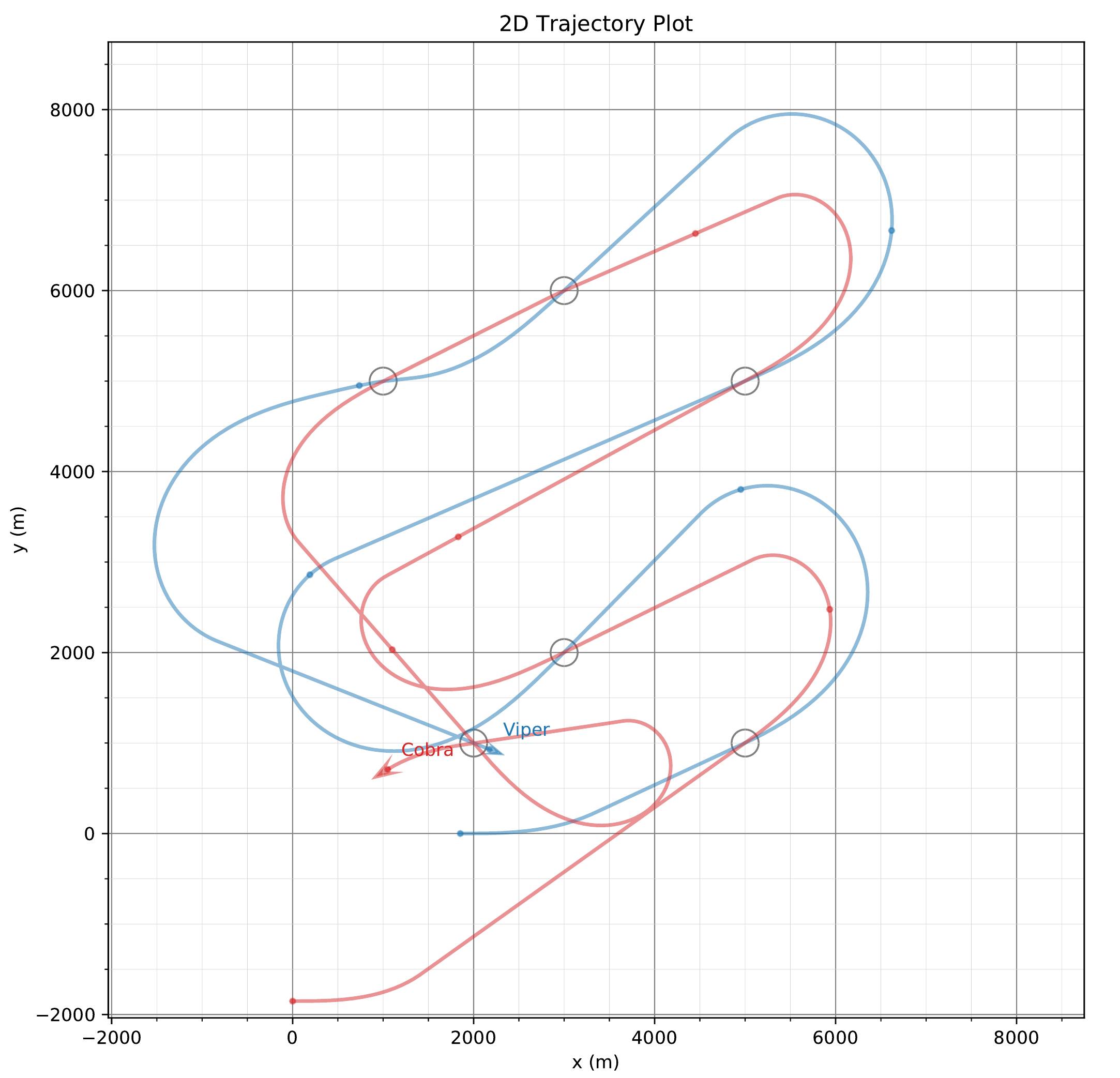}  &
        \includegraphics[width=0.48\textwidth]{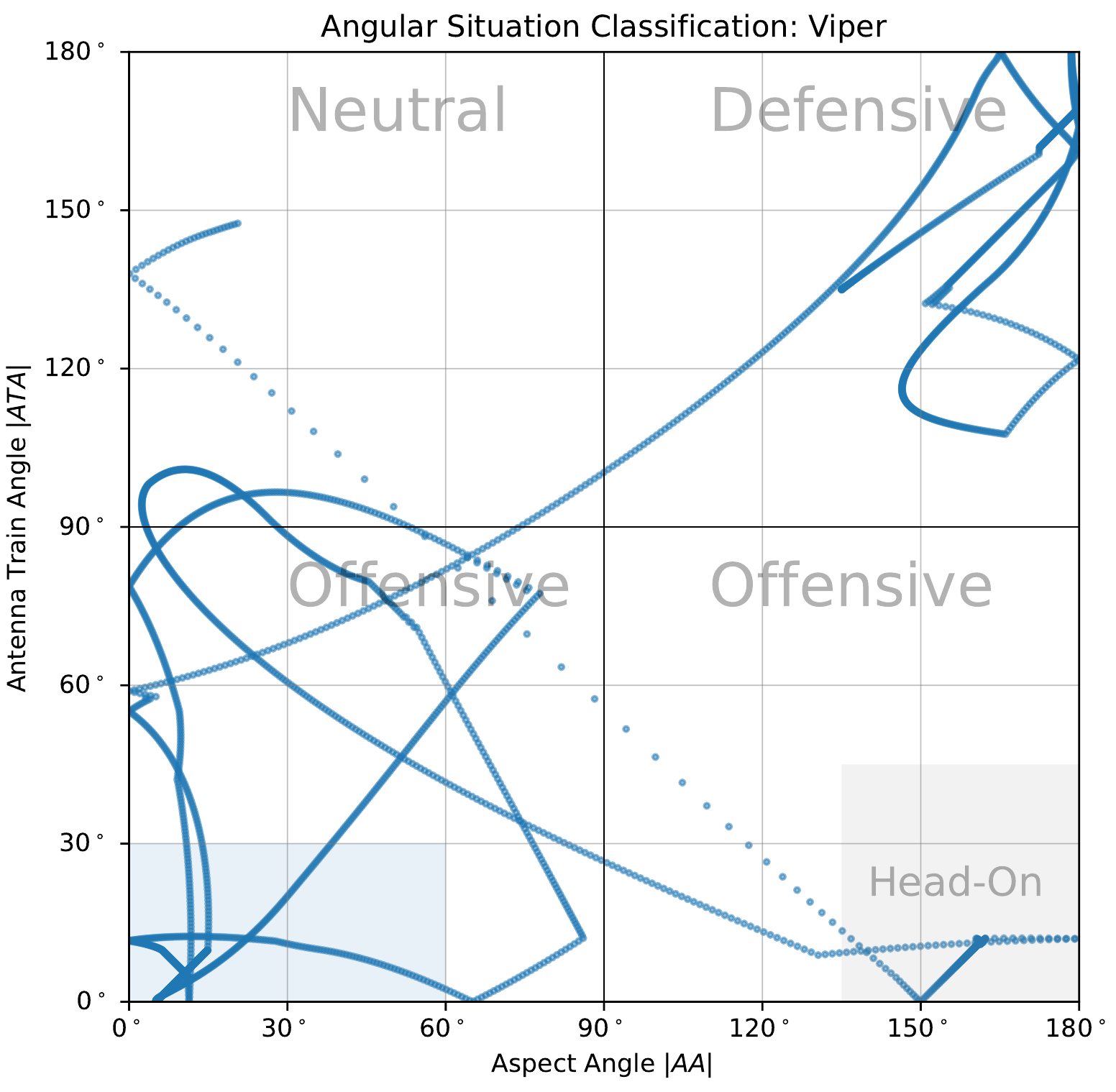} \\
        \small (a) & \small (b) \\
    \end{tabular} 
    \caption{(a) Trajectories of blue and red UAVs from \acezero{} simulation with different starting positions, but
    flying throught the same set of waypoints. (b) The trajectory of the blue UAV relative to the red UAV
    through the situation \emph{orientation space}.  }
    \label{fig:example} 
\end{figure*}

Given a random starting position and orientation, the initial goal of the blue aircraft is to manoeuvre itself behind the red aircraft's tail such that $ATA=0$ and $AA=0$. There may be additional constraints on range and velocity differential given the nature of the blue aircraft's mission. In order to achieve the goal we need to find a policy, behaviour or tactic that satisfies these conditions either as a single manoeuvre or a sequence of
manoeuvres. By plotting the absolute
value of the aspect angle $|AA|$ against the absolute value of the $|ATA|$, we can plot the the trajectory of the blue
aircraft through \emph{orientation space} as demonstrated by Park~\cite{2016:Park}. 
This allows us to classify the angular situation at any given time into broad
categories such as \emph{Offensive}, \emph{Defensive}, \emph{Neutral} and \emph{Head-On}. An example of such a
trajectory can be see in Figure~\ref{fig:example}(b).  

We can use the angular situation information together with other parameters to allow us to score and assess how well the blue \emph{UAV} is doing relative to the \emph{red} UAV. By defining a number of scoring functions we can compare the performance of different AI aerial manoeuvring algorithms.  
We define three scoring functions $S_1$, $S_2$ and $S_3$; some from existing approaches in the literature and some adapted from operational metrics. 

We denote the first scoring function $S_1$ as the \emph{Offensive Quadrant} score as it provides a score of $+1$ if the blue
UAV is located in the bottom left quadrant of the angular situation chart shown in Figure~\ref{fig:example}(b). 
\begin{equation}
    S_1 =
    \begin{cases}
        1, & |AA| \leq \frac{\pi}{2} \: \mathrm{and} \: |ATA| \leq \frac{\pi}{2} \\
        0, & \mathrm{otherwise} 
    \end{cases} 
\end{equation} 

The second scoring function consists of an angular and a range component. The score for the intercepting UAV is
maximised when the $ATA = 0$, $AA = 0$ and the range between the two UAVs is $R = R_d$, where $R_d$ is the desired
range and will be mission dependent. 
The hyperparameter $k$ modulates the relative effect of the range component on the overall score. 
The angular component of the score has been used widely~\cite{1988:Burgin:NASA,1987:Austin}, but the range dependence
was introduced by McGrew~\cite{2010:McGrew} and hence we refer to $S_2$ as the \emph{McGrew Score}. 

\begin{equation}
    S_2 = \frac{1}{2} \left[ \left(1 - \frac{AA}{\pi} \right) + 
                             \left(1 - \frac{ATA}{\pi} \right) 
                      \right] 
          exp \left( \frac{-|R-R_d|}{\pi k} \right)
\end{equation} 

The third scoring function is constructed from Shaw's description~\cite{1985:Shaw} based on the conditions required for a rear quarter
weapon employment against a hostile aircraft. The conditions that must be met for a period of time include constraints
on the aspect and antenna train angles ($|AA| \leq 60^\circ$ and $|ATA| \leq 30^\circ$), a range between the minimum and
maximum range of the weapon system, and a difference in speed less than a nominated minimum. 
This is a strict set of constraints that need to be met and in the terminology of reinforcement learning may be
considered a sparse reward.
We designate this as the \emph{Shaw Score} and define it as follows.

\begin{equation}
    S_3 = 
    \begin{cases} 
        1, &  |AA| \leq \frac{\pi}{3} \mathrm{, \;}  
             |ATA| \leq \frac{\pi}{6} \mathrm{, \;} 
             R_{min} \leq R \leq R_{max} \mathrm{, \;} 
             \Delta v \leq v_{min} \\
        0, & \mathrm{otherwise} 
    \end{cases} 
\end{equation} 

While these scoring functions are not the only ones that can be considered, having a variety of ways of evaluating the
performance of an AI algorithm or approach is important as different approaches will perform differently against
different metrics.  
The scoring functions in practice serve two purposes. First, they are domain-specific operational metrics that allow one AI algorithm to be compared to another. Second, they can be used within an AI algorithm either in their current form or in a modified form to find a solution. For example, they may take the form of a reward function in reinforcement
learning, or a fitness function in an evolutionary algorithm.

\begin{figure}[ht]
    \centering
    \includegraphics[width=0.48\textwidth]{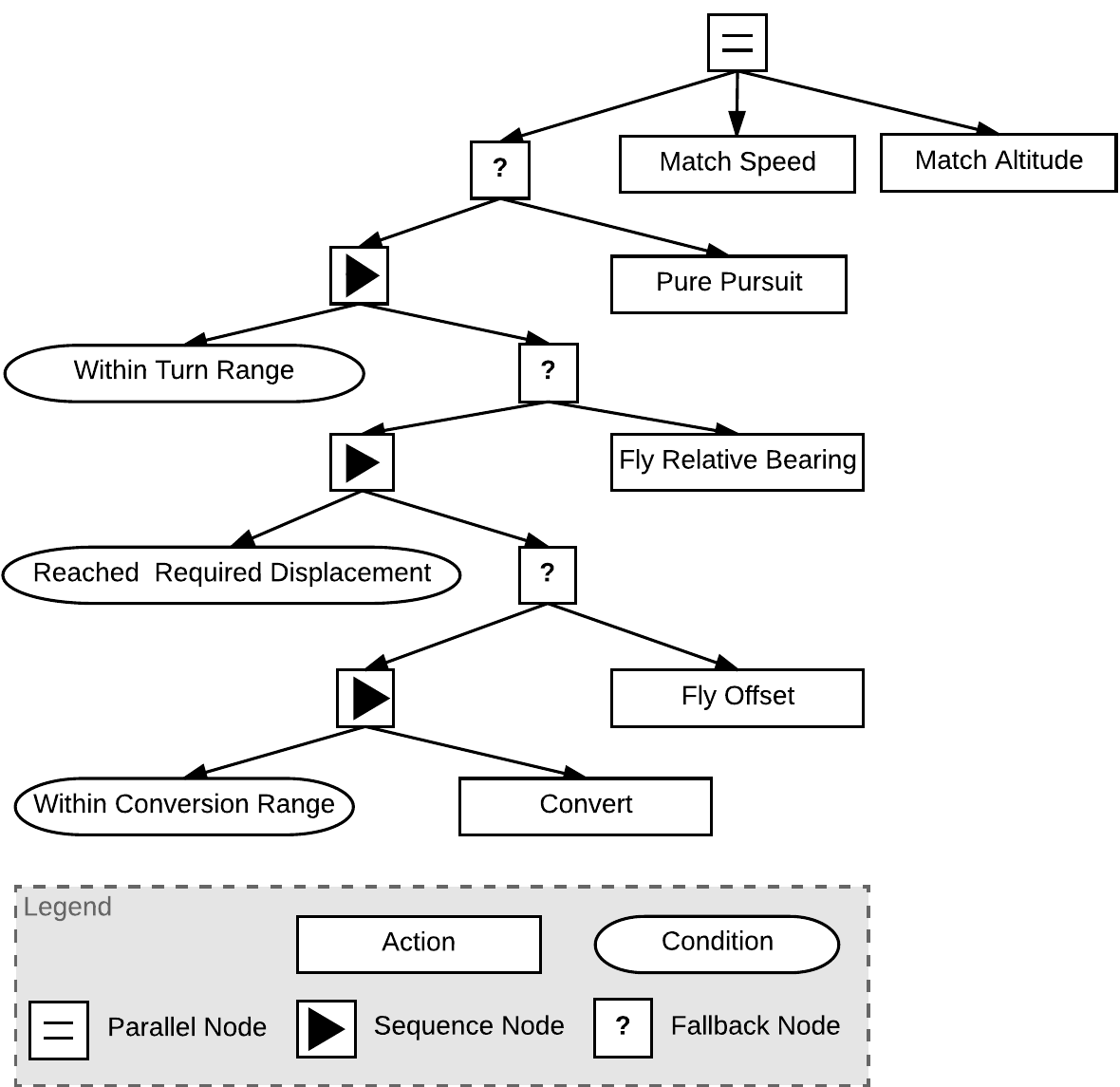} 
    \includegraphics[width=0.48\textwidth]{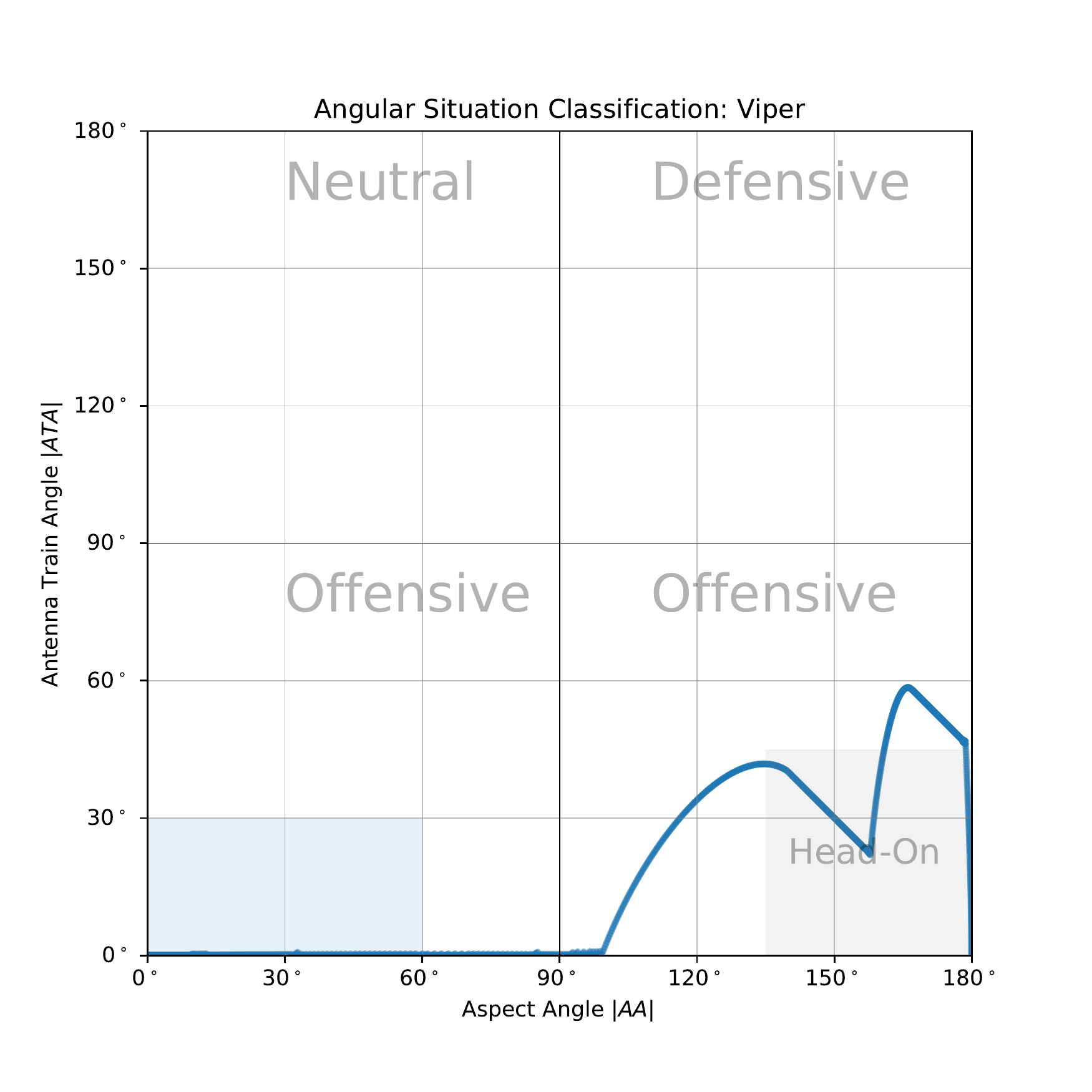}
    \caption{(Left) Implementation of the \emph{Stern Conversion Intercept} as a Behaviour Tree in \acezero{}. Higher-level behaviours such as \emph{PurePursuit} become available as first order actions to exploratory AI algorithms.(Right) Trajectory of a blue aircraft in orientation space successfully performing a \emph{Stern Conversion Intercept}.}
    \label{fig:bt} 
\end{figure}

The scoring functions were implemented in \acezero{} together with a number of baseline handcrafted agents using approaches such as finite state machines and behaviour trees. The environment state space was implemented at multiple levels including both raw state space information (such as the positions, orientations and velocities of each aircraft), as well as engineered features as described previously. The action space was also implemented in a number of ways. Traditional approaches to action space representation in these types of simulations are quite low-level; for example \emph{$\langle TurnLeft, TurnRight, SpeedUp, SlowDown \rangle $}. By implementing baseline behaviours as finite state machines and behaviour trees, the possibility for high-level action spaces becomes available to the algorithms being considered. For example, the behaviour nodes shown in the hand-engineered behaviour tree in Figure~\ref{fig:bt} become available to exploratory algorithms for reasoning at higher levels of abstraction.

\section{Evaluation} 

We provide a qualitative evaluation of the \acezero{} MABS focusing on limitations, the user experience of AI researchers and software engineers and we briefly describe four research projects in which acezero{} was used to evaluate AI behaviour discovery methods in collaboraiton with university partners.

\subsection{Multi-Agent-Based-Simulation Architecture Limitations}
While \acezero{} was specifically designed for exploring AI problems in the aerospace domain, it has provision to support maritime and land units. These can be used support joint domain modelling but their completion is planned for future development. The architecture can be extended to explore AI agent models in other domains (e.g. in games) but this will require additional work that is not available out of the box. 

The system implements a time-stepped simulation architecture that allows variable time-stepped execution on a per component basis but as of this stage does not support event-based simulation.
The architecture was designed with computational models of lower fidelity in mind. Some AI algorithms may require higher fidelity underlying representational models. These are planned to be added on as needed basis driven by the requirements of the research program.

A basic model of agent teaming is supported, where agents can be optionally grouped together to work in teams with a separate agent acting as a team leader or commander. However at this stage more sophisticated models of roles within teams and complex command and control structures are slated for future releases. 

\subsection{User Evaluation}
\begin{figure}
    \centering
    \includegraphics[width=1.0\textwidth]{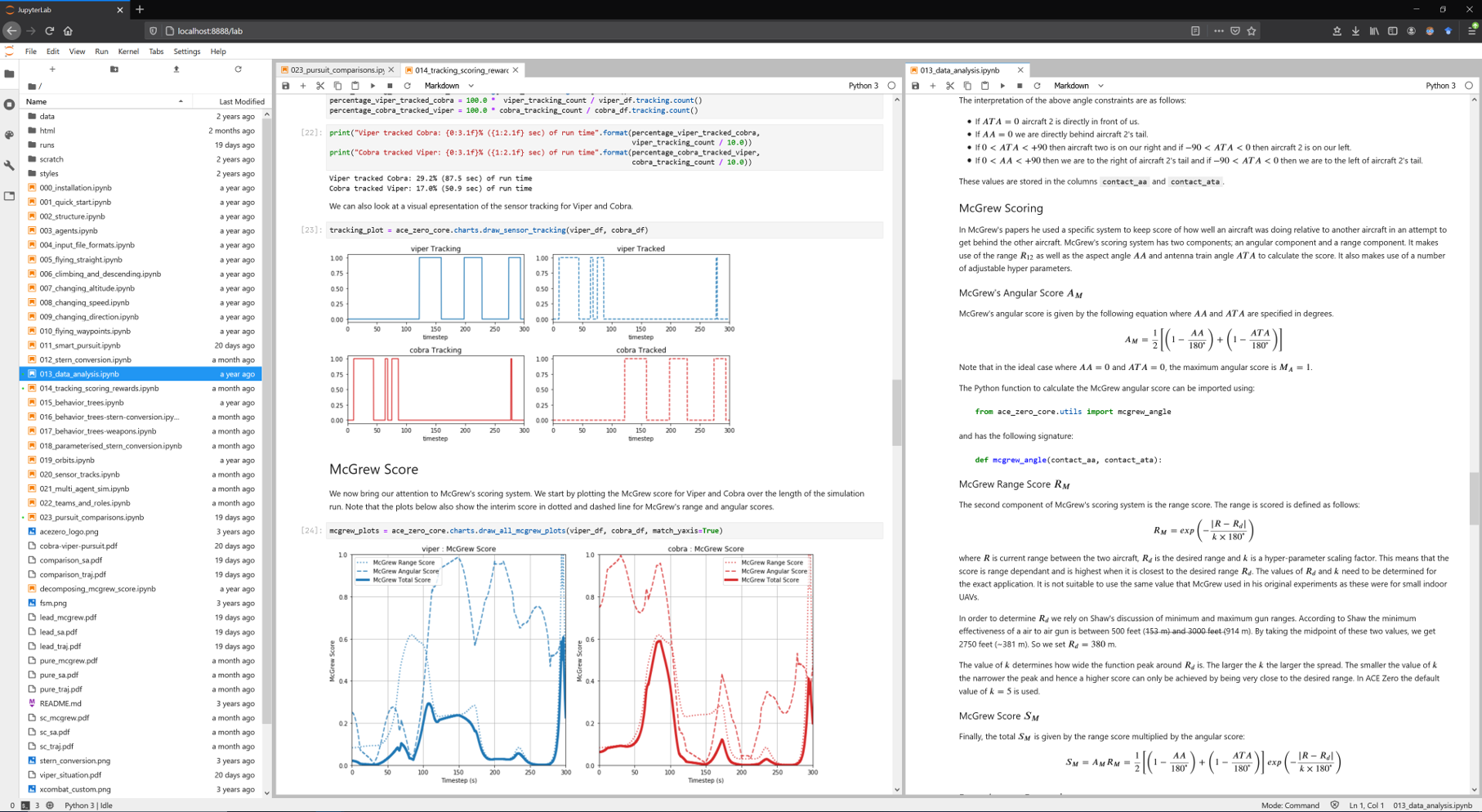}
    \caption{Screenshot of interactive \acezero{} tutorials in the form of computational JupyterLab notebooks.}
    \label{fig:jupyter} 
\end{figure}

The \acezero{} environment was developed for users with software development experience in either computer science or software engineering with a background in artificial intelligence and/or multi-agent systems but not necesarily with any prior knowledge about the aerospace operations domain. 

A key driver was to enable researchers to learn both the simulation environment and the domain relatively quickly so that they could focus their efforts in their specific area of agent decision making expertise. To do this, a series of interactive computational tutorials in the form of JupyterLab notebooks~\cite{2007:Perez:IPython,2016:Kluyver:JupyterNotebooks} were developed. The notebooks not only formed the core documentation for \acezero{} but also allowed researchers to experiment with parameters and agent behavioural models and observe the results in real-time. Figure~\ref{fig:jupyter} shows a screenshot of the tutorial interface.
The researchers were also provided with a copy of XCombat a 3D animation tool for visualising the trajectories of the aircraft flown by the pilot agents (see Figure~\ref{fig:xcombat}). 

Over twenty notebooks were developed covering a spectrum of topics ranging from installation, to explaining the architecture, details about the aerospace domain, explaining the agent percept and action interfaces, metrics and measures of effectiveness for evaluating, data processing, analysis and visualisation. Notebooks were also used as a step by step guide on rapidly developing pilot agent models using either finite state machines (FSMs) or Behaviour Trees (BTs), that could be deployed immediately and run within the JupyterLab notebook environment. 

The tutorials start with describing how to develop a pilot agent to control the simulated aircraft using lower level actions such as turning, climbing, descending and changing speed. These basic building blocks are then used to develop more sophisticated maneuvers, initially as an individual aircraft, then relative to another adversarial aircraft and finally developing more complex tactics for multiple aircraft working together as team. These examples are followed by tutorials on the calculation of metrics for evaluating agent performance as well as data analysis and visualisation for explaining agent behaviour.  

To date, over twenty five researchers from six institutions (four universities, a public R\&D lab and a software development company) have used \acezero{} to develop agents for behaviour discovery as will be described in Section~\ref{sec:research-applications}. This has included researchers and engineers with a range of expertise including professors, lecturers, post-doctoral researchers, software engineers and PhD, masters and undergraduate students. By all accounts the combination of the tutorials as well as ongoing support and collaboration from the development team allowed the researchers to get up to speed with the environment in a relatively short time. In most cases the researchers were able to go through all the tutorials in one or two days, by the end of which they had each built simple agents working in \acezero{} and had an initial introduction to the domain. This compares very favourably with current production level multi-agent simulation environments which have a steep learning curve and typically require a two week training course (a one week analyst's course and a one week software developer's course) to get started.

\subsection{Research Applications}
\label{sec:research-applications}

\subsubsection{Evolutionary Algorithms} 
The first investigation into exploratory AI and behaviour discovery algorithms considered evolutionary
algorithms~\cite{2018:Masek:PRIMA,2019:Lam:ORP}. 
The first phase of the project started with taking an existing FSM implementation of
the stern conversion intercept and evolving the tactical parameters to result in an optimally evolved tactic. This was followed 
by taking the basic behaviours in an FSM, deleting the transitions between them and evolving a new behavioural agent
with evolved state transitions. The third phase involved breaking the behaviours down even further and looking at evolving
behaviours from the low level commands available to the UAV. More complex behaviours in the form of behaviour trees
(BTs) were explored in the fourth and fifth stages of the project. 
The research investigated the effect of evolving tactical behaviours for
complex behaviour trees and using a library of conditions and behaviours, and then using genetic programming methods to generate
new behaviour trees~\cite{2021:Masek:GP}. Developing a viable cost function was a primary challenge in all the evolutionary algorithm research undertaken.

\subsubsection{Automated Planning} 
The second project investigated the application of automated planning using width-based search techniques~\cite{2012:Lipovetzky} to develop an agent capable of
executing a stern conversion manoeuvre. In order to generate a plan for the \emph{blue} aircraft one must be able to predict the state of the simulation (for blue and red) at a finite horizon. Since classical planning is a model-based approach, a model of
the dynamics of the system was required. As such this work involved a novel combination of hybrid planning with optimal
control~\cite{2017:Ramirez:ijcai,2018:Ramirez:aamas} (specifically model predictive control) that resulted in a high-performing pilot agent capable of manoeuvring the \emph{blue} UAV to achieve the goal.
As a follow-on from this research project, \acezero{} was used to
investigate behaviour recognition using planning, building on the methods developed by Vered~\cite{2017:Vered:IJCAI}. 

\subsubsection{Generative Adversarial Networks} 
The third project involved exploring the feasibility of Generative Adversarial Networks (GANs) for behaviour generation
in \acezero{}. 
The aim of this work was to generate new tactical behaviour based on examples of existing \emph{successful} behaviours or
tactics. The initial focus in this project was to consider behaviours (or plans) as sequences of
goal-directed actions. As such, a generalised technique to generate goal-optimised sequences could be used not only for
tactical behaviour but also other sequences (such as text generation~\cite{2018:Hossam:TextGen}). 
While the research here is still in
progress, further details can be found in the work on OptiGAN~\cite{2020:Hossam:OptiGAN} that uses trajectory data
from \acezero{}, with a combination of a GAN and a reinforcement learning (RL) approach to generate 
sequences of actions to achieve the goal of a stern conversion manoeuvre. 

\subsubsection{Reinforcement Learning} 
The fourth project (currently in progress) is focused on generating pilot tactical behaviour (i.e. policies) using reinforcement learning. The project has two distinct parts. The first part is to look at traditional reinforcement learning techniques with a view to exploring multi-objective reinforcement learning (MORL) using \acezero{}. Preliminary results have investigated reward structures~\cite{2019:Kurniawan:Rewards} and supervised policy learning~\cite{2020:Kurniawan:D2DSL}. 

The longer term plan is to explore deep reinforcement learning in \acezero{}. Specific research questions of interest include exploring state space representations, continuous action spaces, reward shaping and more importantly, multi-agent and team based reinforcement learning to discover and learn new tactical behaviours for teams of autonomous aircraft. 

\section{Conclusions} 
The \acezero{} MABS has undergone an initial round of evaluation in four separate academic research groups with specialties in different sub-fields of artificial intelligence. The outcomes from these initial studies are currently being evaluated for consideration to transition into a large scale simulation environment.
\acezero{} is intended as a lightweight environment for AI algorithm evaluation, and as a result may not capture all of the nuances that arise in a large-scale deployed simulation environment. A number of challenges arise when developing an environment such as this. The first is maintaining a balance between simplicity and the complexity present in a production simulation environment. The second is one of scale; while some behaviour discovery approaches might be viable in a simpler environment, the extended computation time required by a more complex environment might make the approach unviable for practical use. 

The results from the previously described research projects provide a baseline for considering which AI techniques should be transitioned from research into development for production-level simulation environments. In addition to assessing the suitability and viability of a specific method for operations analysis studies, the experience also allows the operations researchers to scope resource requirements in terms of budget, schedule and staffing (i.e. AI and software engineering expertise required for deployment).

A number a future challenges remain. First, a deeper understanding must be developed of the suitable levels of abstraction for state space, action space and cost/reward representation. Second, in a future where algorithms can discover new tactical behaviour, how do we measure both the novelty and robustness of the discovered behaviour? Finally, we need to consider how exploratory AI methods can be integrated into existing techniques and methods in agent-oriented software engineering.


\bibliographystyle{splncs04}
\bibliography{papasimeon,refs,software}

\end{document}